\documentclass[%
aps,prb,
longbibliography,
amsmath,amssymb,
superscriptaddress,
onecolumn,
12pt,
floatfix%
]{revtex4-2}

\usepackage{graphicx} 
\usepackage{dcolumn} 
\usepackage{bm} 

\usepackage[utf8]{inputenc}
\usepackage[T1]{fontenc}
\usepackage{mathptmx}
\usepackage{amsmath}
\usepackage{xcolor}
\usepackage[colorlinks,citecolor=blue,linkcolor=blue]{hyperref}
\usepackage{soul}
\usepackage[normalem]{ulem}

\usepackage{tikz,xcolor}
\definecolor{lime}{HTML}{A6CE39}
\DeclareRobustCommand{\orcidicon}{%
	\begin{tikzpicture}
	\draw[lime, fill=lime] (0,0) 
	circle [radius=0.16] 
	node[white] {{\fontfamily{qag}\selectfont \tiny ID}};
	\draw[white, fill=white] (-0.0625,0.095) 
	circle [radius=0.007];
	\end{tikzpicture}
	\hspace{-2mm}
}
\foreach \x in {A, ..., Z}{%
	\expandafter\xdef\csname orcid\x\endcsname{\noexpand\href{https://orcid.org/\csname orcidauthor\x\endcsname}{\noexpand\orcidicon}}
}

\newcommand{\unit}[1]{\,\mathrm{#1}}

\begin{document}

\title{Quantum Memories for Fundamental Science in Space}

\author{Jan-Michael Mol\orcidA{}}
\email{jan-michael.mol@dlr.de}
\affiliation{Deutsches Zentrum für Luft- und Raumfahrt e.V. (DLR), Institute of Quantum Technologies, Ulm, Germany}

\author{Luisa Esguerra}
\affiliation{Deutsches Zentrum für Luft- und Raumfahrt e.V. (DLR), Institute of Optical Sensor Systems, Berlin, Germany}
\affiliation{Technische Universität Berlin, Institut für Optik und Atomare Physik, Berlin, Germany}

\author{Matthias Meister\orcidC{}}
\affiliation{Deutsches Zentrum für Luft- und Raumfahrt e.V. (DLR), Institute of Quantum Technologies, Ulm, Germany}

\author{David Edward Bruschi\orcidD{}}
\affiliation{Institute for Quantum Computing Analytics (PGI-12), Forschungszentrum J\"ulich, 52425 J\"ulich, Germany}

\author{Andreas Wolfgang Schell\orcidE{}}
\affiliation{Institut f\"ur Festk\"orperphysik, Leibniz Universit\"at Hannover, 30167 Hannover, Germany}

\author{Janik Wolters\orcidF{}}
\affiliation{Deutsches Zentrum für Luft- und Raumfahrt e.V. (DLR), Institute of Optical Sensor Systems, Berlin, Germany}
\affiliation{Technische Universität Berlin, Institut für Optik und Atomare Physik, Berlin, Germany}

\author{Lisa W\"orner\orcidG{}}
\email{lisa.woerner@dlr.de}
\affiliation{Deutsches Zentrum für Luft- und Raumfahrt e.V. (DLR), Institute of Quantum Technologies, Ulm, Germany}

\date{\today}

\begin{abstract}
Investigating and verifying the connections between the foundations of quantum mechanics and general relativity will require extremely sensitive quantum experiments. To provide ultimate insight into this fascinating area of physics, the realization of dedicated experiments in space will sooner or later become a necessity. Quantum technologies, and among them quantum memories in particular, are providing novel approaches to reach conclusive experimental results due to their advanced state of development backed by decades of progress. Storing quantum states for prolonged time will make it possible to study Bell tests on astronomical baselines, to increase measurement precision for investigations of gravitational effects on quantum systems, or enable distributed networks of quantum sensors and clocks.  We here promote the case of exploiting quantum memories for fundamental physics in space, and discuss both distinct experiments as well as potential quantum memory platforms and their performance.
\end{abstract}

\maketitle

\newpage

\section{Introduction}
Quantum technologies are currently expanding into viable public and commercial applications as well as extending their capabilities for use in engineering and applied science. There is increasing interest to deploy such technologies in space to advance secure quantum communication~\cite{GisinThew2007,SangouardGisin2011}, assist distributed quantum computation~\cite{KimbleKimble2008,WehnerHanson2018,PirandolaBraunstein2016}, improve sensing~\cite{DegenCappellaro2017}, and run experimental tests of fundamental physics~\cite{BelenchiaBassi2022,MohagegKwiat2021}. Operation of complex quantum systems in space has already been successfully demonstrated by the Cold Atom Laboratory~\cite{AvelineThompson2020} and MAIUS~\cite{BeckerRasel2018} missions, operating Bose-Einstein condensates (BECs) in space. The upcoming project BECCAL~\cite{FryeWoerner2021} will build on this heritage and perform a multitude of new experiments, among them the operation of a quantum memory in space. Since quantum memories are an important component for future quantum communication systems, insights from this mission will inform further development of space-ready hardware for global quantum networks.

In general, quantum memories allow to store a given quantum state, such as the state of a single photon, for a specific amount of time until retrieval occurs at a later point in time. The key aspect here is that this process of storage and retrieval conserves all previously established quantum properties, e.g. entanglement and quantum coherence. In this way, quantum memories are able to play a key role in greatly enhancing many scientific and technological applications, such as quantum state teleportation and long-range Bell tests~\cite{PirandolaBraunstein2015,UrsinZeilinger2009,RenPan2017,LinHu2021,CaoPan2018} and aiding the design of new quantum tests of general relativity~\cite{BruschiFuentes2014,BruschiRazavi2014,JoshiUrsin2018,BruschiSchell2021a,BruschiSchell2021}.

\begin{figure}[t!]
\centering
\includegraphics[width=.9\textwidth]{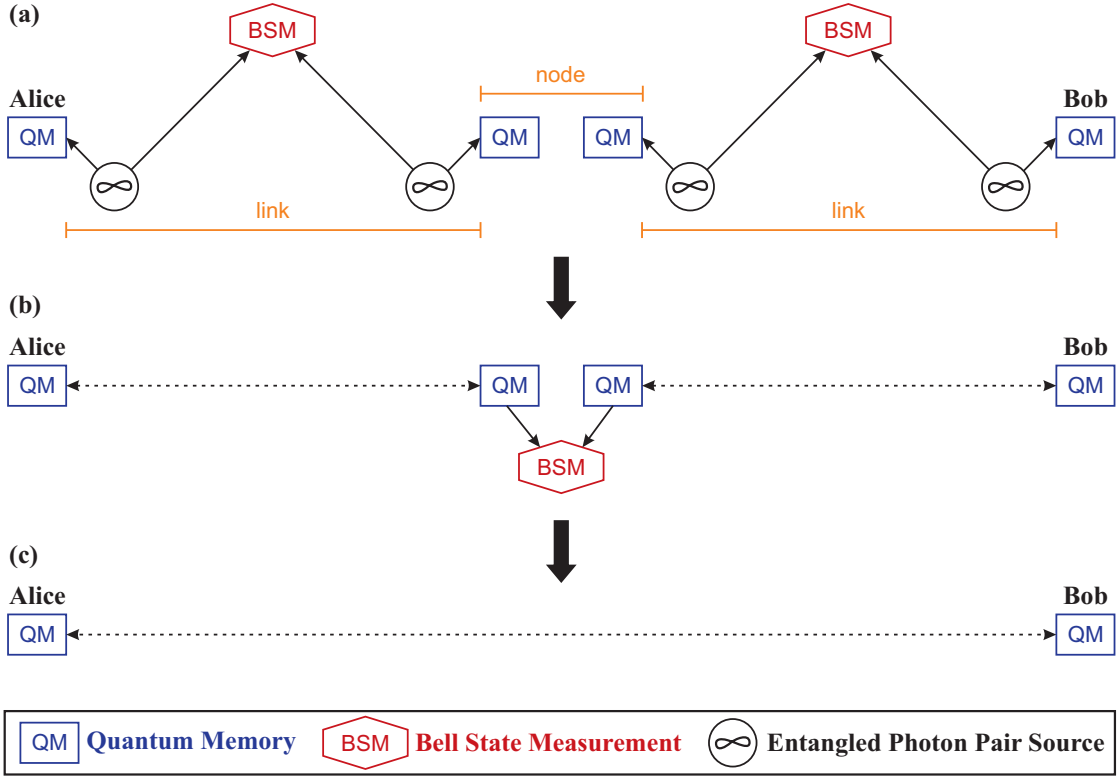}
\caption{Single-node quantum repeater network. \textbf{(a)} In a first step, Alice and Bob each entangle their local quantum memories with a quantum memory at a repeater node. \textbf{(b)} Once both entangled links are established, another Bell state measurement between the two node memories generates the entangled link between the communicating parties in \textbf{(c)}. In this scheme, memories play a crucial role in synchronization of entangled link generation. More nodes may be concatenated to increase the number of links and thus bridge larger and larger distances.}
\label{fig:repeater}
\end{figure}

We therefore highlight the necessity of space-borne quantum memories for fundamental quantum physics experiments over long distances and timescales. Even though we are primarily interested in quantum memories with long coherence times for this matter, large scale quantum networks and long distance quantum communication will require quantum repeaters which make use of quantum memories as a sub-component (see Fig.~\ref{fig:repeater}). The necessary distribution of entanglement between two parties is based on Bell-state-measurements (BSM) of incident signals at connecting nodes. To bridge large distances, repeaters can consequently be utilized to combat losses and increase signal range. A repeater necessarily requires elements to receive and transmit signals, as well as to perform BSMs on the signals themselves. In this configuration, quantum memories are deployed to synchronize two incoming signals by individually storing photons for a desired period of time. In principle, repeaters can be linearly connected in-between two communicating parties to bridge larger and larger distances, subject only to the overall channel fidelity. The requirements for useful memories in quantum repeaters go far above the capabilities of simple fiber-delay-loops which cannot offer on-demand retrieval and furthermore suffer from significant inherent losses for longer storage times, i.e. increasing fiber length. For longer storage and on-demand retrievability, other types of memories can be chosen, as will be discussed in Section~\ref{sec:qumemtypes}. Ultimately, the desired storage time depends strongly on the application. For instance, measuring gravitational effects potentially requires longer coherence times than quantum secure information transmission (Section~\ref{sec:meminspace}).

Combining quantum memories and free-space photon links in space helps to overcome several limitations of ground-only experiments by mitigating loss from long optical fiber links or atmosphere~\cite{ValloneVilloresi2015,ScriminichVedovato2021,SidhuOi2021}, extending limited line-of-sight, or providing a low-disturbance microgravity environment~\cite{LachmannRasel2021}. Optical losses could in principle also be bridged by memory-based quantum repeaters on Earth, but the required number of repeaters poses a substantial challenge~\cite{MuralidharanJiang2016,SidhuOi2021}. To illustrate the advantage of space operation, the Quantum Experiment at Space Scale aboard the Micius satellite already showed order of magnitude improvement at extending the range of quantum entanglement distribution compared to fiber-based approaches~\cite{YinPan2017}. In its mission, flight hardware carried entangled photon sources, with detection and measurement remaining on-ground. Hence, it can be concluded that employing quantum memories at memory-assisted ground nodes in conjunction with multiple satellites is a next step to increase the range of entanglement distribution even further. Finally, intermediary nodes may be moved onto satellites, improving fidelity by avoiding unnecessary light paths through the atmosphere altogether~\cite{GuendoganKrutzik2021,WallnoeferWolters2021}. Building space-based quantum networks in this manner not only aids the establishment of future global quantum communication, but may also be utilized in fundamental science studies in space. 

\begin{figure}[t!]
    \centering
    \includegraphics[width=0.9\linewidth]{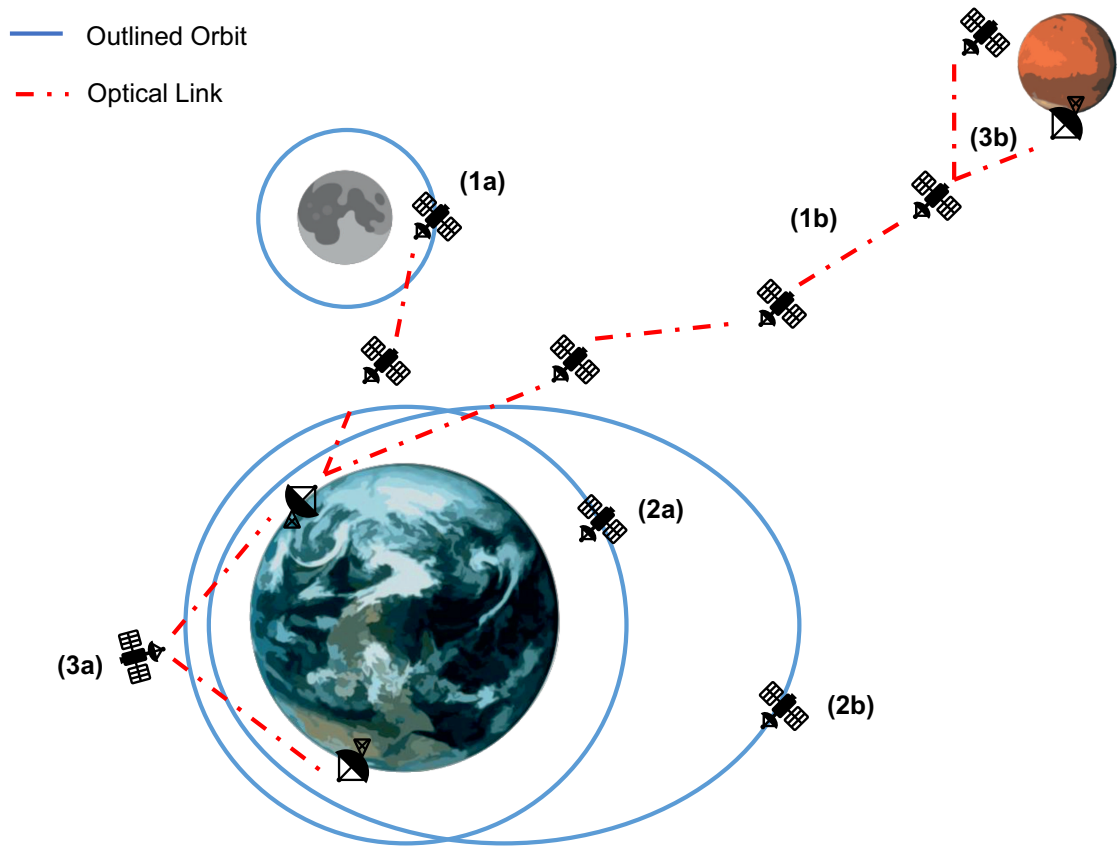}
    \caption{Exemplary applications for quantum memories in space. \textbf{Case 1} describes long-range Bell tests in which quantum memories are deployed either as part of satellite-based repeaters for bridging large distances or by storing photons locally to fulfill space-like separation requirements (Section~\ref{subsec:bell}). Depicted are measurements involving ground stations, transport vehicles and an orbiter surrounding the Moon \textbf{(1a)} or Mars \textbf{(1b)}. \textbf{Case 2} illustrates the collection of phases due to gravitational gradients. An entangled photon is sent to a satellite and stored in on-board memory, while its partner is stored on Earth. \textbf{(2a)} and \textbf{(2b)} show different conceivable orbits for such experiments which may be used for comparative measurements (Section~\ref{subsec:localcurved} \& ~\ref{subsec:entdyn}). \textbf{Case 3} is a description of quantum information transfer between two distant stations. Those could either be placed on Earth \textbf{(3a)} or on another celestial body \textbf{(3b)}. Both cases have been described elsewhere and require a network of quantum repeaters to enable global coverage and long distance transmission~\cite{GuendoganKrutzik2021}). However, an exemplary network bridge to Mars will likely require more than ten thousand operating nodes with current technology. This poses a major challenge which will need to be addressed for deep space applications.}
    \label{fig:network}
\end{figure}

In Section~\ref{sec:meminspace}, we therefore want to give an outlook on how space-borne quantum memories can improve various experiments in space. A depiction of several scenarios is shown in Figure~\ref{fig:network}. It outlines long-range Bell tests (Sec.~\ref{subsec:bell}) and measurements of gravitational or motional effects (Secs.~\ref{subsec:localcurved} and \ref{subsec:entdyn}). Once larger networks become available, tasks such as more precise time-keeping (Sec.~\ref{subsec:clkprec}) and distributed sensing (Sec.~\ref{subsec:distribsens}) become feasible.

Section~\ref{sec:qumemtypes} will cover available options as well as requirements and limitations of different quantum memory platforms, where we representatively discuss warm vapor cells, cold atomic gases, rare-earth-ion-doped crystals, color centers in diamond, and single atoms and ions in cavities. We conclude the section by giving a summary of engineering challenges lying ahead.

\section{Memory-Assisted Fundamental Science in Space} \label{sec:meminspace}
Space-based quantum science is generating an increasing amount of interest. In the past years there have been many proposals for potential missions to study a wide variety of quantum phenomena in space \cite{JoshiUrsin2018,MagnaniVinjanampathy2019,VillarLing2020,MohagegKwiat2021}. As we will see in the following subsections, there are proposals that require entanglement between distant users to be maintained for long duration in contrast to those that primarily require a quantum state to be stored securely, but not necessarily for longer time periods.

\subsection{Long-range Bell Tests} \label{subsec:bell}
The Copenhagen interpretation of quantum mechanics leads to the conclusion that a theory of nature cannot both be local and realistic. This was mathematically expressed by John Stewart Bell through an inequality that states an upper bound for correlations between measurement outcomes of distant particles which obey both realism and locality. Numerous quantum physical experiments have shown the violation of this inequality~\cite{HensenHanson2015,GiustinaZeilinger2015,ScheidlZeilinger2010,ShalmNam2015}, with some experiments even ruling out detection and locality loopholes. However, to achieve statistical significance in these experiments averaging of temporally subsequent measurements creates a memory loophole, which can only statistically but not fundamentally be ruled out. A longer baseline could help in creating a true space-like separation between all measurements and thus close this loophole as well.

While quantum memories are not a necessary requirement to run Bell tests over longer distances, they are able to aid such tests by storing part of an entangled quantum state until measurements can be synchronously carried out in satisfaction of space-like separation requirements. A simple setup could comprise an entangled photon-pair source in which one photon is locally stored in a quantum memory for delayed measurement, while the other is sent off to a remote location in the meantime. This type of configuration would avoid the need for a second distributional baseline and thus reduces the requirement to only one single long baseline.
Leaving loopholes aside, to demonstrate the technical feasibility of this approach one could install a ground-based entangled photon-pair source and a quantum memory which sends off one photon for detection to a satellite in low-Earth-orbit (LEO), located $\sim200-2000$~km above ground. The required coherence time for such a memory lies in the range of milliseconds, which has already been realized in current memories~\cite{BaoPan2012,JobezAfzelius2015}.

Moving further ahead, source and memory might be placed on a LEO-satellite or the International Space Station (ISS), and establish a link to a detector orbiting the Moon. This enables Bell tests investigating ``Freedom-of-Choice'' loopholes by letting two experimenters on Moon and Earth determine their measurement basis during the travel time of the photon \cite{MohagegKwiat2021}. The required memory coherence in this case increases to above one second, which poses a challenge for current memories at single-photon level, but storage times of several hours for many-photon states~\cite{ZhongSellars2015,MaGuo2021} raise hopes to reach this goal in the near-future.

\subsection{Effects of curved spacetime on localized quantum systems} \label{subsec:localcurved}

\begin{figure}[t!]
    \centering
    \includegraphics[width=\textwidth]{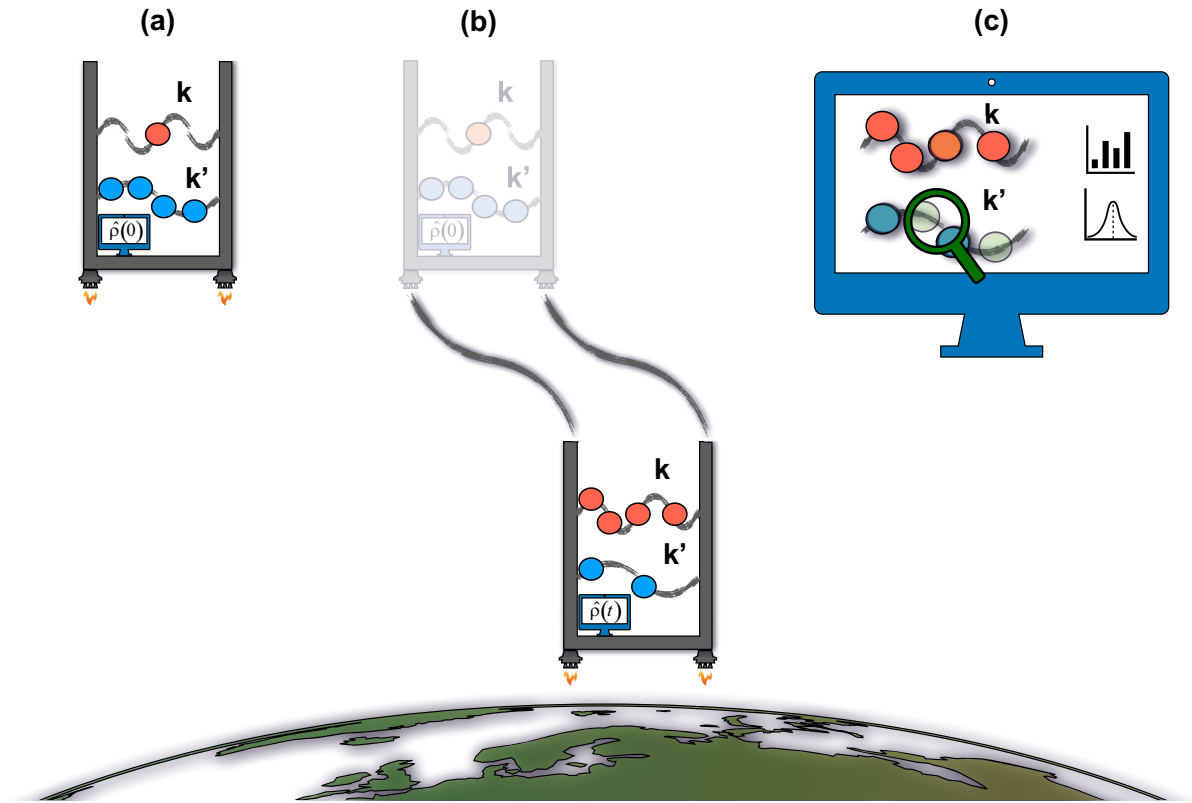}
    \caption{\textbf{(a)} A quantum memory is placed in space. The initial state of the degrees of freedom that encode the desired information is $\hat{\rho}(0)$. \textbf{(b)} The spacecraft containing the memory is moved for some time $t$, and the state of the quantum system is altered. \textbf{(c)} In general, the state $\hat{\rho}(t)$ of the quantum systems that do constitute the memory -- say the modes $k$  and $k'$ of a quantum field -- is (slightly) entangled with other degrees of freedom, thereby affecting the ability to faithfully retrieve the quantum state initially stored. Quantum metrology techniques are then used to obtain bounds on the precision of measurements of relevant parameters.}
    \label{fig:entanglement:motion}
\end{figure}

Improving precision in measurements allows for the development of novel technologies as well as for the discovery of new physical effects. In the context of interest to this work, the quantities to be measured will be most likely encoded in degrees of freedom placed at different locations, or nodes, or the resources to be exploited for the task will be distributed across space~\cite{BruschiFuentes2014}. Generally speaking, modes of a quantum field or other degrees of freedom will be entangled either within the same site,  across sites, or both. One or more of the nodes, holding a quantum memory where the information or state is stored, can undergo a physical process that will affect the entanglement shared. This can occur at the detriment of the precision, due to effective loss of information to unwanted or not accessible degrees of freedom, but it can also enhance the overall precision if the changes are engineered correctly. Changes to the whole setup might include motion of one or more of the crafts holding the memories~\cite{BruschiFuentes2014} or a strong dynamical change in the local spacetime curvature~\cite{SabinFuentes2014}. The language to deal with obtaining the bound of the achievable precision of measurements that can be performed in these cases is provided by the field of \textit{quantum metrology}.

Quantum metrology empowers (quantum) technologies to potentially beat the state of the art by taking advantage of genuine quantum features, such as entanglement~\cite{GiovannettiMaccone2006}. In this context, a unitary evolution $\hat{U}(\vartheta)$ transforms an initial state $\hat{\rho}(0)$ and encodes a parameter $\vartheta$ of interest in the final one. This could be, for example, the action of a gravitational wave affecting the quantum state stored inside a quantum memory in space. Concretely, we have $\hat{\rho}(0)\rightarrow\hat{\rho}(\vartheta)=\hat{U}(\vartheta)\hat{\rho}(0)\hat{U}^\dag(\vartheta)$. The lower bound on the precision $\Delta\vartheta$ that can be achieved is given by the Cram\'er-Rao bound $\Delta\vartheta\geq(N \mathcal{H})^{1/2}$, where $N$ is the number of measurements or input individual quantum states, and $\mathcal{H}$ is the quantum Fisher information (QFI)~\cite{RaoRao1992}. This important information-theoretical quantity is related to (theoretical) operational ways to distinguish nearby quantum states in the Hilbert space~\cite{BraunsteinCaves1994}. Measurements on quantum states are routinely performed in the laboratory today. However, there are many new effects that exist on timescales where it is virtually impossible to maintain the quantum coherence of quantum systems. This gap can be bridged by quantum memories, which could allow for maintaining of the state long enough for the effects to occur. 

One case of particular interest is that of dynamical creation of entanglement within moving cavities. It has been shown that (relativistic) motion of a cavity that confines a quantum field can not only create particles, the well-known Dynamical Casimir effect (DCE)~\cite{DodonovDodonov2010}, but it can also create quantum correlations between the field excitations~\cite{BruschiLouko2012,FriisLouko2012,FriisFuentes2012,BruschiLouko2013}. A systematic study of the effects of motion unveiled the properties of the transformations involved, and how they can be engineered to produce different (entangling) quantum gates~\cite{FriisBruschi2012,BruschiFuentes2013}. Specific techniques were subsequently developed to apply the concepts of quantum metrology to a relativistic setting, and an ideal application was proposed for gravitational wave sensing using entangled states of phonons within a BEC~\cite{SabinFuentes2014}. In these studies the cavities were assumed to be infinitely rigid. While this is a good approximation for proof-of-principle studies, elastic properties of the cavity must be taken into account. This can be done, and preliminary attempts in this direction in the same context of the aforementioned studies have been successfully conducted~\cite{RaetzelFuentes2018,BravoFuentes2022}. More work is necessary to properly characterize the system including these aspects for concrete mission proposals.

Precise measurements based on relativistic quantum fields are therefore a potentially promising avenue, which is nevertheless marred by the ability to sustain quantum states for a long time for later processing. The key issue here is, therefore, the ability to store the quantum states created, while the cavity is located in an isolated environment, thereby reducing mechanical vibrations and other constraints due to being Earth-bound. The state of the field, once stored, can then be affected by the motion of the support that confines the memory and this, in turn, results in the entanglement initially present in the state to be affected as well. It has been shown that specifically designed resonances enhance the effects (on the entanglement or the average particle number) linearly or quadratically with time \cite{BruschiFuentes2013}. A rough estimate shows that longer times are desirable in photonic cavities within the considered perturbative regimes. The question is different, however, if the characteristic frequencies inherent in the particular memory are smaller than those of light. In that case, it was shown that perturbations of the phononic perturbations of a supporting many-body system (such as the phonons of a BEC) are affected by gravity and motion, and therefore the techniques discussed here apply as well. This means that quantum memories with characteristic small frequencies can witness larger effects in a shorter time.
A depiction of such scenarios can be found in Figure~\ref{fig:entanglement:motion}.

\subsection{Entanglement dynamics due to motion}\label{subsec:entdyn}
\begin{figure}[t!]
    \includegraphics[width=\textwidth]{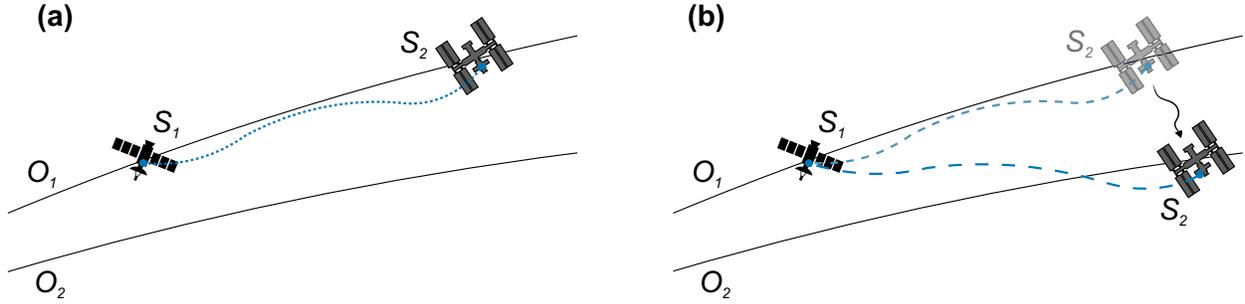}
    \caption{\textbf{(a)} Two satellites $S_1$ and $S_2$ travel on the same orbital path $O_1$. This makes them both inertial. Two memories hosted inside the satellites are entangled with each other (symbolized by the dotted line). \textbf{(b)} Satellite $S_2$ performs a manoeuvre that brings it onto orbit $O_2$, where it will be inertial again. The process of orbit transfer inevitably requires satellite $S_2$ to be non-inertial for a period of time. During this process, the entanglement between the memories will be altered due to the effects of non-inertial motion on the confined quantum systems, as depicted in Figure~\ref{fig:entanglement:motion}(a,b).}
    \label{fig:entanglement:degradation}
\end{figure}

Quantum memories can also be used to boost sensitivity of entanglement-related measurements. For these applications, the storage of quantum states for prolonged time is required to probe effects in changing gravitational environments~\cite{RideoutTerno2012}. For example, one proposal puts forward the use of two entangled cold atomic gases to probe gravitation- or acceleration-induced decoherence~\cite{BruschiFuentes2014}. By use of two memories located on two different spacecrafts, an initial joint quantum state entangles both memories when they are located on the same orbit corresponding to a shared inertial frame. Afterwards, one of the spacecrafts is accelerated to a different orbit. As discussed in~\ref{subsec:localcurved}, it has been shown that the motion of a cavity affects the quantum states of localized systems confined within, such as the field modes of quantum field \cite{BruschiLouko2012} or the phononic modes of a BEC \cite{SabinFuentes2014}. Therefore, it is expected that any entanglement shared between the degrees the modes in the moving cavity and other quantum systems will be affected as well. Estimates of necessary coherence times range from 100 ms~\cite{BruschiFuentes2014} to seconds or minutes using more conservative assumptions for reasonable sensitivity. Multiplexing capabilities to generate high enough signal-to-noise through statistical data are likely necessary here, depending on memory efficiency and fidelity.
The scheme is depicted in Figure~\ref{fig:entanglement:degradation}.

\subsection{Quantum memories for improved clock precision} \label{subsec:clkprec}
Precise time-keeping is not only interesting for standardization~\cite{GuinotArias2005}, but also positioning accuracy in applications such as global navigation satellite systems, or possibly mapping of gravitational potentials~\cite{ChouWineland2010}. Sharing timing information within a classical network of clocks can already help to decrease the Allan deviation of involved nodes by classical averaging. A quantum network of $N$ clocks distributed across a finite volume in space has been proposed in order to increase time keeping precision even further~\cite{KomarLukin2014}. By utilizing quantum entanglement, clock nodes are potentially able to interrogate and stabilize the average frequency of the network via a shared quantum state, instead of classically operating and correcting each node separately. For short averaging times, this leads to a $\sqrt{N}$ improvement in precision of fully-quantum networks compared to those basing their operation solely on classical interrogation and cooperation.

Entanglement propagation between clocks can either be achieved by direct exchange of flying qubits or by using quantum repeaters. While timing issues in entanglement generation can be solved by smart design of the synchronisation algorithm, quantum memories may be able to offer more flexiblity by storing states until clock qubits are ready for further processing, especially in larger and more complex networks. In case all generated photon pairs for entanglement distribution need to be simultaneously stored in one memory, the optimal number of modes for a clock scales as $[\mathrm{log}(n)]^2$~\cite{KomarLukin2014}, requiring a mode capacity of 85 per channel and clock containing $n=10^4$ qubits. This is well in reach for present quantum memories~\cite{SeriRiedmatten2019,PuDuan2017,LipkaParniak2021}.

To provide a global time reference, a sufficiently large clock network necessarily has to operate in space, due to the already discussed losses associated with ground-only optical links.

\subsection{Distributed quantum sensing} \label{subsec:distribsens}
Similar to the quantum-cooperative clock network described above, distributed networks of quantum sensors can utilize the same underlying principle to enable operation of distributed sensing, where noise scales with $1/N$ for a fully quantum-operated network compared to $1/\sqrt{N}$ for a fully classically-operated one, giving rise to a $\sqrt{N}$ improvement. Large enough networks may therefore lead to sensitive measurements below the standard quantum limit~\cite{ZhangZhuang2021,ZhuangShapiro2018,GuoAndersen2020} and enable detection of gravitational effects, possibly linking spacetime to quantum physics~\cite{RideoutTerno2012}.

One particular proposal for using distributed quantum sensors is their application in astronomy to build interferometric telescopes~\cite{GottesmanCroke2012} or optical wavelength telescope arrays~\cite{KhabiboullineLukin2019,KhabiboullineLukin2019a}. Stationary Earth-bound quantum memories will likely be the first realizable generation of hardware and are needed to support any quantum sensor network. Again, the evolution towards space-based arrays of quantum sensors and memories as envisioned in quantum communication networks will yield similar benefits for astronomy and geodesy. The feasibility of applying quantum networks to geodesic measurements has recently been shown by detecting and localizing seismic events~\cite{ChenPan2022}.

\section{Quantum Memory Types} \label{sec:qumemtypes}
\begin{figure}[t!]
    \centering
    \includegraphics[width=0.6\textwidth]{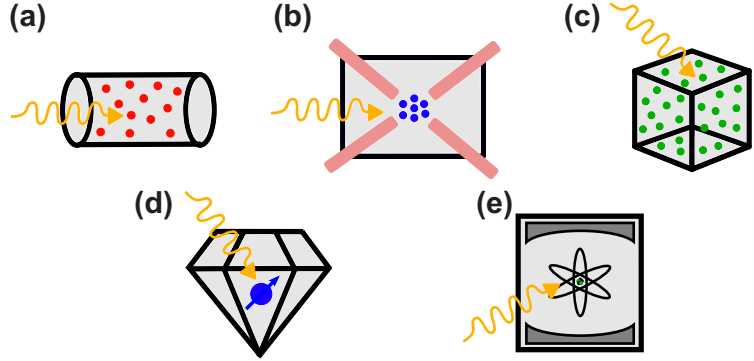}
    \caption{Possible quantum memory platforms. Among ensemble-based photon storage, we find \textbf{(a)} warm vapor cells, \textbf{(b)} cold atomic gases in magneto-optical traps, and \textbf{(c)} rare-earth-ion-doped crystals (REIDs). Using single emitters, noticeable examples are \textbf{(d)} color centers in diamond or \textbf{(e)} single atoms/ions in a cavity.}
    \label{fig:memory_types}
\end{figure}

To enable the experiments discussed in Section~\ref{sec:meminspace}, reliable storage of photons (flying qubits) for a flexible or predefined amount of time has to be achieved. Quantum memories have the capability of acting as interfaces between flying and stationary qubits. They are able to absorb and re-emit photonic qubits on-demand or after a specific time defined by the system itself. Many different platforms exist for the realization of quantum memories (see Fig.~\ref{fig:memory_types}). One can differentiate between ensemble-based platforms, such as cold and warm atomic gases, as well as rare-earth-ion-doped crystals (REIDs), and single emitters, such as nitrogen-vacancy (NV) centers in diamond or single atoms and molecules. Optically active systems serve as prime candidates due to already advanced technology for photonic distribution of quantum states. We therefore focus on these platforms and do not discuss those which currently lack optical interfaces, despite some of them featuring exceptional coherence times and fidelities~\cite{StegerPohl2012}. 

A variety of protocols exists to implement quantum storage for different platforms. They can be divided into optically controlled ones, for example electromagnetically induced transparency (EIT) \cite{PhillipsLukin2001}, Raman-type schemes \cite{ReimWalmsley2010}, the fast-ladder memory (FLAME) \cite{FinkelsteinFirstenberg2018}, and the off-resonant-cascaded-absorption (ORCA) memory \cite{KaczmarekWalmsley2018}, as well as engineered absorption schemes, where one finds gradient-echo memory (GEM) \cite{HosseiniBuchler2011}, controlled-reversible-inhomogeneous broadening (CRIB) \cite{KrausCirac2006}, and atomic-frequency comb (AFC) \cite{AfzeliusGisin2009,AfzeliusKroell2010,MainLedingham2021} protocols. Each platform often supports more than one protocol. Proper selection highly depends on the planned memory application. So far, no single combination yields a sufficient solution for a general-purpose device, i.e. trade-offs exist between large efficiencies, long storage times, temporal and spatial multimode capacity, large bandwidths, high fidelities, etc. In case of memory efficiencies, special attention has to be paid to not only include memory-intrinsic values, but also of the setup as a whole. In the end, these trade-offs have to be balanced with respect to a targeted application. As an example, on-demand readout may not be necessary in scenarios where storage times are already fixed beforehand, such as in the case of Bell tests over (approximately) constant baselines. This may relax requirements on memory protocols, which in turn can reduce system complexity, for example by circumventing the need for long-term spin-wave storage in rare-earth-ion-doped crystals, if optical coherence times suffice already.

In the following subsections, a selection of the most relevant platforms will be described in more detail, and exemplary state-of-the-art implementations will be mentioned. Some of the experiments have been performed storing bright laser pulses containing many photons instead of operating at single-photon level. However, true single photons are mandatory for most scenarios. Therefore, the memories need to be paired with single-photon sources, which poses an additional challenge on the memory platforms~\cite{HeshamiSussman2016}, as well as on the development of efficient photon sources~\cite{MottolaWolters2020} matched to the memories.

\subsection{Warm vapor cells} \label{subsec:warmvapcells}
Vapor cells are low-complexity systems which can be operated over a wide range of temperatures and conditions without the need for any cryogenic refrigeration and large magnetic fields, which makes them an easily scalable platform. Commonly used are alkaline atoms since they possess energetically low-lying spin states and long coherence times. These states function as storage states in EIT and Raman protocols~\cite{HeshamiSussman2016}. Atomic vapors show high optical depths even at room temperature and thus high storage efficiencies can be achieved. However, these types of memories are known to be susceptible to noise, especially from fourwave-mixing~\cite{MichelbergerWalmsley2015}. 

By use of the spin-orientation degrees of freedom of Cs atoms and a special chamber coating, Katz and Firstenberg~\cite{KatzFirstenberg2018} report memory efficiencies of 9--14\% at storage times up to 150 milliseconds. By applying a specific week-long temperature cycle to combat spin destruction, they are able to increase storage times up to 430 milliseconds and claim that the memory is sufficient to store weak coherent states and squeezed states. Through the hyperfine manifold, they posit possible single photon operation. Guo et al.~\cite{GuoZhang2019} achieve the highest efficiency to date in warm vapor memories with a value of 82\% by an off-resonant Raman scheme in a Rubidium memory. They further obtain fidelities at single photon level up to 98\%. Focusing on the interfacing of the memories with a single photon source, in~\cite{WoltersTreutlein2017} a memory with an acceptance bandwidth of 0.66 GHz is realized, which is suitable for photons from semiconductor quantum dots. 
Low noise levels in atomic vapors are achieved in~\cite{ThomasLedingham2019}, who report on values of $\mu_1=0.20$ noise photons without performing cavity engineering schemes as done in~\cite{SaundersNunn2016}. In~\cite{NamaziFigueroa2017} noise levels below $\mu_1=0.04$ are reported.

All of the above results are achieved as a trade-off to remaining memory parameters. A simultaneous optimization of two relevant parameters, the end-to-end efficiency and signal-to-noise ratio at single photon level is performed in~\cite{EsguerraWolters2022}. True single photon storage and retrieval in a ground-state warm atomic vapor memory using a photon from a spontaneous-parametric-down-conversion source is demonstrated for the first time in~\cite{BuserTreutlein2022}.

Spatial multiplexing for multimode capacity is feasible in these systems, at the cost of more powerful lasers~\cite{MessnerWolters2021}, but has to our knowledge not yet been realized experimentally. Microfabrication techniques have enabled cell dimensions of millimeter size, which may aid miniaturization for satellite integration~\cite{KitchingKitching2018}.

\subsection{Cold atomic gases} \label{subsec:coldatomicgases}
In contrast to vapor cells, cold atomic gases are cooled in magneto-optical traps to reduce atomic motion to 100 $\mu$K or below which provides a route for reaching long coherence times. The creation of BECs at even lower temperatures requires either stronger lasers for deeper traps or atom-chips for RF-evaporative cooling. Atom-chip assemblies have aided miniaturization~\cite{KeilFolman2016} and manipulation, but require more careful planning of vacuum feedthroughs and interaction between atom cloud and chip surface. 

Cold gases generally implement the same memory protocols as warm gases. High efficiencies and storage times have been achieved; in particular,~\cite{YangPan2016} reports an intrinsic retrieval rate of 76\% for storage times of 220 milliseconds in a Rubidium gas. In~\cite{BaoPan2012} memory efficiencies of 73\% have been achieved, although at lower storage times of 3.2 milliseconds. Multimode capability has been realized through orbital angular momentum~\cite{DingGuo2013} or spatial multiplexing~\cite{PuDuan2017}, and entanglement between two Rubidium memories has recently been demonstrated~\cite{YuPan2020}.

Portable versions of cold atomic gases already exist~\cite{BeckerRasel2018,SalimAnderson2011,FarkasAnderson2010} and on-chip optics might in future alleviate the need for bulkier optical components~\cite{StraatsmaSalim2015}. Nevertheless, the upper limit of photon absorption rate puts a constraint on minimum system size for reasonable BEC creation in the centimeters range. Especially interesting for space applications are developments in miniaturization of passively pumped systems with 1000-day UHV operation using microfabricated magneto-optical traps~\cite{RushtonHimsworth2014}. 

Heritage from previous missions MAIUS~\cite{BeckerRasel2018} and CAL~\cite{AvelineThompson2020} make cold atom gases an important contender for space quantum memories, with project BECCAL~\cite{FryeWoerner2021} serving as a future cornerstone.

\subsection{Rare-earth-ion-doped crystals (REID)} \label{subsec:reids}
The inherent strengths of solid-state based systems lie in their low-complexity, compactness, and micro-integration possibilities. In REIDs, optical modes down to single-photon level are stored in a collective excitation of the dopant rare-earth ions in a crystal which is typically cooled to temperatures below 6 K. Prominent dopants are Praseodymium ($\mathrm{^{141}Pr^{3+}}$) or Europium ($\mathrm{^{151/153}Eu^{3+}}$) embedded in an Yttrium-Orthosilicate matrix ($\mathrm{Y_2SiO_5}$), although various other materials exist~\cite{WelinskiGoldner2020,AhlefeldtSellars2016,AskaraniTittel2021}, even compatible with telecommunication wavelengths~\cite{LauritzenGisin2010,CraiciuFaraon2019}.

Optical coherence times range can approach $100\unit{\mu s}$~\cite{AskaraniTittel2021,CraiciuFaraon2019,HorvathRippe2021,OrtuAfzelius2018}, in some cases even hundreds of $\unit{\mu s}$~\cite{BoettgerSun2009,LimMorton2018,WelinskiGoldner2020}. To achieve longest coherence times, optical excitations can be converted into spin waves whose dephasing can then be suppressed by using dynamical decoupling techniques~\cite{HolzaepfelAfzelius2020}. Systems based on europium ($\mathrm{^{151}Eu}$) donors furthermore enable operation at a magnetically insensitive transition around 1 Tesla, achieving spin coherence times ranging up to six hours~\cite{ZhongSellars2015} and optical storage up to one hour~\cite{MaGuo2021}. 
Single-photon operation has been achieved with photon noise levels of $\mu_1 = 0.069(2)$~\cite{GuendoganRiedmatten2015} and $\mu_1 = 0.10(2)$~\cite{JobezAfzelius2015} for comparable storage times around $12\unit{\mu s}$.

Most REIDs are operated as hybrid-AFC memories to store a frequency comb in a spin-wave for on-demand retrieval, although EIT memories exist as well~\cite{SchraftHalfmann2016,HeinzeHalfmann2013}. Down-conversion to match telecom wavelength is possible~\cite{SeriRiedmatten2017,SeriRiedmatten2019} which was recently used for multimode operation between two REID memories~\cite{LagoRiveraRiedmatten2021} or to interface with cold atomic gases~\cite{MaringRiedmatten2017}.

When considering system miniaturization for space flight, the requirement of bulky magnets can be circumvented by operating at smaller but sufficient bias fields~\cite{OrtuAfzelius2022,HolzaepfelAfzelius2020}. Crystals can also be integrated using optical waveguides~\cite{CorrielliDeRiedmatten2016}. Recent developments in compact cryocoolers for single photon detectors benefit cooling requirements of solid-state based systems as well~\cite{YouXie2018}.

\subsection{Color centers in diamond} \label{subsec:colorcenters}
Other contenders for a solid-state quantum memory are the various color centers in diamond or silicon carbide~\cite{SonAwschalom2020}. Well studied is the nitrogen vacancy (NV) center in diamond, where a substitutional nitrogen atom next to a vacancy inside the carbon lattice of bulk diamond leads to magnetically tunable fluorescence. In addition to its use as a sensitive magnetometer, the color center has general access to the quantum properties of the associated free electron spin as well as the nitrogen atom and any proximal $^{13}$C nuclei. These nuclei have recently been utilized to form a ten-qubit memory register with coherence times of 75 seconds for arbitrary single-qubit states and more than 10 seconds for two-qubit entanglement~\cite{BradleyTaminiau2019}. Techniques to improve storage times include entanglement distillation via $\mathrm{^{13}C}$ nuclei~\cite{ReisererMarkham2016}, isotopic purification of diamond to minimize inhomogeneous broadening~\cite{BalasubramanianWrachtrup2009}, and strain engineering~\cite{SohnLoncar2018}.

Although NVs are currently the most mature system, which have recently been complemented by tin-vacancy (SnV)~\cite{DebrouxAtatuere2021} or silicon-vacancy (SiV)~\cite{SukachevLukin2017} centers, which offer better compatibility in photonic nanostructures due to their inversion symmetry, yielding high collection efficiencies~\cite{BhaskarLukin2020}. In comparison to the low temperatures needed for longer coherence in SiV, similar or better coherence at temperatures above 1 K might be offered by SnV. 

Color centers are mostly read out using confocal microscopes which impact system size considerably. Optical cavities~\cite{BhaskarLukin2020}, photonic chips~\cite{MachielseLoncar2019,WanEnglund2020,WoltersBenson2010} or photoelectric readout~\cite{SiyushevJelezko2019} are therefore helpful developments to shrink system sizes. Additionally, since color centers also serve as excellent single-photon emitters ~\cite{SipahigilLukin2014}, they are able to provide native sources in integrated systems.

Low efficiencies in absorptive storage schemes may make vacancy centers more attractive as emissive quantum memories, which can be of use in Bell tests (Sec.~\ref{subsec:bell}) or distributed clocks and sensors (Secs.~\ref{subsec:clkprec} and~\ref{subsec:distribsens}).

\subsection{Single atoms and ions} \label{subsec:atomsandions}
Single atoms and ions in high finesse cavities were among the first candidates investigated for quantum memories~\cite{KimbleKimble2008,SangouardSimon2009,ReisererRempe2015}. They offer excellent coherence properties, combined with efficient optical interfaces. Tasks like entanglement swapping have been demonstrated~\cite{OlmschenkMonroe2009}. Using high finesse cavities, the optical interaction can be enhanced by the Purcell effect and even strong coupling to a cavity mode can be achieved~\cite{KimbleKimble1998,WilkRempe2007,JungeRauschenbeutel2013,TakahashiKeller2020}. 

Used as memories, single ions allowed for about 70~\% efficiency with~\cite{SchuppLanyon2021}, limited by the ion's level structure. Deterministic entanglement at a fidelity of 90.1(17)~\% between a trapped Yb ion and a photon emitted into the resonator mode was achieved~\cite{KobelKoehl2021}. A memory based on a single atom reached 22~\% efficiency and a storage time of 100~ms \cite{KoerberRempe2018}. Moreover, single atoms allow for heralded storage \cite{KalbRempe2015} and when coupling to nearby particles \cite{WilkBrowaeys2010,CiracZoller1995} basic quantum information processing can be realized, which is useful for quantum error correction.

However, the experimental complexity is high and with the limited multimode capacity imposes a hard limit to applications that require multimode memories, e.g. for entanglement distribution.

\subsection{Engineering challenges for quantum memories in space} \label{subsec:engchallenges}
Our discussion highlights the special interest to mount quantum memories to space borne platforms. As space-based research and technology development always underlies restrictions with respect to size, mass, and power, investigations to integrate quantum technologies into CubeSats~\cite{OiUrsin2017,VillarLing2020} or satellites are a natural development~\cite{TakenakaToyoshima2017}. This goes alongside hardening against harsh environmental requirements, as required for a lunar mission or beyond~\cite{MohagegKwiat2021,MazzarellaWoerner2021}. 

With long distances being a major concern in transferring single photons to execute quantum measurements, the feasibility demonstration of quantum-limited signal propagation from geostationary Earth orbit (GEO) to ground by Alphasat I-XL~\cite{GuenthnerLeuchs2017} is a major milestone. This underlines that memories in geostationary orbits or as relay stations are a viable and necessary option to bridge long distances. Successful feasibility studies of Bell-tests over high-loss channels also paint a positive picture for space-based fundamental science~\cite{CaoPan2018}.

\begin{table}[tb!]
\caption{\label{tab:cmp}Exemplary quantum memories and their key performance indices. Efficiencies are typically memory intrinsic. "Store \& Retrieve" indicates whether photon storage and/or retrieval are possible.}
\begin{ruledtabular}
\begin{tabular}{l|l|l|l|l|l}
  \textbf{Platform} & \textbf{Mechanism} & \multicolumn{1}{p{2cm}|}{\textbf{(S)tore \& \newline (R)etrieve}} & \textbf{Bandwidth} & \textbf{Efficiency} & \multicolumn{1}{p{2cm}}{\textbf{Storage \newline Time}}\\ \hline
  \textit{Warm vapor cells} & & & & & \\
    \quad $^{87}$Rb~\cite{GuoZhang2019} & Raman & S \& R & 77 MHz & 82.0\% & 170 ns \\
    \quad $^{87}$Rb~\cite{WoltersTreutlein2017} & EIT & S \& R & 660 MHz & 17\% & 50 ns \\ \hline
  \textit{Cold atomic gases} & & & & & \\
    \quad $^{87}$Rb~\cite{BaoPan2012} & EIT & R & - & 73\% & 3.2 ms \\
    \quad $^{87}$Rb~\cite{YangPan2016} & EIT & R & - & 76\% & 220 ms \\ \hline
  \textit{Rare-earth-ion-doped crystals} & & & & & \\
  \quad $\mathrm{^{151}Eu^{3+}:Y_2SiO_5}$~\cite{OrtuAfzelius2022} & hybrid-AFC & S \& R & 1.5 MHz & 7.4\% & 20 ms \\
    \quad $\mathrm{^{141}Pr^{3+}:Y_2SiO_5}$~\cite{YangLi2018} & hybrid-AFC & S \& R & < 2 MHz & 5\% & 13 $\mathrm{\mu s}$ \\ \hline
  \textit{Color centers in diamond} & & & & & \\
    \quad NV~\cite{YangWrachtrup2016} & Absorption & S & 12 MHz & 1\% & > 10 s \\
    \quad SiV~\cite{BhaskarLukin2020} & Dispersive & S &$\approx$50 MHz & - & 0.2 ms \\ \hline
  \textit{Single atoms and ions} & & & & & \\
    \quad $\mathrm{^{87}Rb}$~\cite{KoerberRempe2018} & Raman & R & - & 22\% & 100 ms \\
    \quad $\mathrm{^{171}Yb^+}$~\cite{KobelKoehl2021} & Raman & R & 19.4 MHz & 10.1\% & 1.022 ms \\
\end{tabular}
\end{ruledtabular}
\end{table}

Deploying optical hardware in space will inevitably have to take into account the challenges of alignment and robustness of components, such as high numeric aperture lenses and others. Solutions can be found in already existing technology such as laser communication terminals~\cite{SmutnyCzichy2009,HeineBenzi2015, GuenthnerLeuchs2017} and optical clocks~\cite{GiorgiSchuh2019,SchuldtBraxmaier2021}.

When it comes to the preferential choice of quantum memory platform for space missions, multiple factors have to be considered: in the intermediate run, warm vapor cells and cold atomic gases provide the most advanced capabilities for miniaturization due to already existing flight heritage and relatively low-complexity peripherals. Their high optical efficiencies and bandwidth currently put them ahead as well, but longer storage times may ultimately be provided by REIDs and color centers. Relatively compact cryocoolers for satellites are readily available and can offer cooling powers in excess of 100 mW at 35 K. For typical temperature requirements of solid-state quantum memories of 6 K and below, cooling power and associated size, weight and power budget require more engineering effort. Also, vibrational influence on optical assemblies is a major concern, especially in smaller footprint satellites where compressors cannot easily be spatially separated from cold-stages. Nevertheless, solid-state-based systems are rapidly progressing and will likely find applications on-ground in the intermediate run, before more compact cryogenics or higher temperature operation will make them more suitable for space missions in the long-term. Additionally, possible photonic integration promotes them as excellent candidates in the long run, both in system size and memory efficiency.

In the end, key performance factors which are exemplified in table~\ref{tab:cmp} have to be addressed with respect to respective experiments to be executed. The list of entries is by no means exhaustive, but should give a representative overview for applications discussed in this paper. It highlights the various trade-offs to be considered when optimizing for explicit application requirements.

\section{Conclusion}
Space-borne quantum memories have a huge potential in boosting fundamental physics experiments~\cite{JoshiUrsin2018,MagnaniVinjanampathy2019,VillarLing2020,MohagegKwiat2021,BelenchiaBassi2022}. Once the ability to perform long-distance Bell tests has been established, memory-assisted experiments will serve as an excellent platform for more complex measurements of relativistic effects on localized quantum systems that share entanglement between moving observers \cite{BruschiFuentes2014}, as well as experiments located within regions of spacetime with dynamical gravitational fields~\cite{SabinFuentes2014}. Not only will quantum memories allow for tests of predictions of science at the overlap of relativity and quantum mechanics, but these investigations will also support future technological developments for deep-space quantum communications \cite{MazzarellaWoerner2021,MohagegKwiat2021}.

We have discussed ideas that lie behind recent predictions of the effects of motion and dynamical gravitational fields on the quantum state of a localized system, and how testing such predictions requires storage of quantum states for a long time. Storage times and efficiencies in quantum memories are approaching a useful regime for space applications, and multimode capabilities are currently heavily investigated. Additionally, the many activities which are ongoing in the quantum technology and quantum memory communities provide an optimistic outlook for improvements in the next few years.

\acknowledgments
We thank Luca Mazzarella and Albert Roura for insightful discussions about interesting space applications for quantum memories.

\bibliography{bibliography}

\end{document}